\documentclass[twocolumn,showpacs,preprintnumbers,amsmath,amssymb]{revtex4}

\usepackage{graphicx}
\usepackage{dcolumn}
\usepackage{bm}

\begin{document}

\preprint{NTT-qubit/860504-2004-04}

\title{Coherent control of a flux qubit by phase-shifted resonant microwave pulses}
\author{Tatsuya Kutsuzawa$^{1,2,3}$, Hirotaka Tanaka$^{1,3}$, Shiro Saito$^{1,3}$, Hayato Nakano$^{1,3}$, Kouichi Semba$^{1,3}$, and Hideaki Takayanagi$^{1,2,3}$}

\affiliation{$^{1}$NTT Basic Research Laboratories, NTT Corporation, Atsugi, 243-0198, Japan\\
$^{2}$Department of Physics, Tokyo University of Science, 1-3 Kagurazaka, Shinjuku-ku, Tokyo 162-8601, Japan\\
$^{3}$CREST, Japan Science and Technology Agency, Japan}
\date{\today}
             
\begin{abstract}
The quantum state of a flux qubit was successfully pulse-controlled by using a resonant microwave. We observed Ramsey fringes by applying a pair of phase-shifted $\frac{\pi}{2}$ microwave pulses without introducing detuning. With this method, the qubit state can be rotated on an arbitrary axis in the x-y plane of the Bloch sphere in a rotating frame. We obtained a qubit signal from a coherent oscillation with an angular velocity of up to 2$\pi\times$11.4G rad/s. In combination with Rabi pulses, this method enables us to achieve full control of single qubit operation. It also offers the possibility of orders of magnitude increases in the speed of the arbitrary unitary gate operation.
\end{abstract}

\maketitle

 A superconducting circuit containing Josephson junctions is a promising candidate as a quantum bit (qubit), which is an essential building block for future quantum computers \cite{NC-book00}.
 In the superconducting circuit, the qubit is represented by two quantized states which are collective states of a ``macroscopic" number of Cooper pairs.
 Recently, the NMR-like coherent control of several types of these Josephson qubits has been reported \cite{Yasu-Nature99,VACJPUED-Sci02,YCCW-Sci02,MNAU-PRL02,CNHM-Sci03}.
 In addition to Rabi, Ramsey, and spin echo type experiments on a single qubit, conditional gate control with a more complex pulse sequence has been demonstrated for a two qubit system \cite{YPANT-Nature03} or in a system consisting of a qubit interacting with a harmonic oscillator \cite{Delft-Nature04,Yale-Nature04}.
 Although the controllability of these qubits has been rapidly developed, the typical coherence time of the Josephson qubit remains rather short compared with other qubit candidates.
This has made the effective use of this precious coherence time and the realization of a shorter gate operating time desirable.  

 In quantum computation, it is essential to control each qubit by performing arbitrary unitary operations at will.
 For one qubit, Rabi oscillation and Ramsey fringes experiments provide information related to the control of the qubit state $|\Psi\rangle = \cos{\frac{\theta}{2}}|0\rangle+ e^{i\phi} \sin{\frac{\theta}{2}}|1\rangle$. In the Bloch sphere notation, Rabi oscillations give us both information and the ability to control the polar angle $\theta$.
 The Ramsey fringes give us the ability to control the azimuth angle $\phi$.
 However, the observation of the Ramsey fringes of a flux-qubit usually involves  a few hundred MHz detuning from the qubit resonant frequency. 
 In this paper, we propose a new method for observing Ramsey fringes, the phase shift method, which can control the phase of microwave (MW) pulses at the resonant frequency of the qubit. The advantage of this method is that it provides nearly two orders of magnitude faster azimuth angle $\phi$ control of a qubit than the conventional detuning method.

\begin{figure}
\includegraphics[width=8cm]{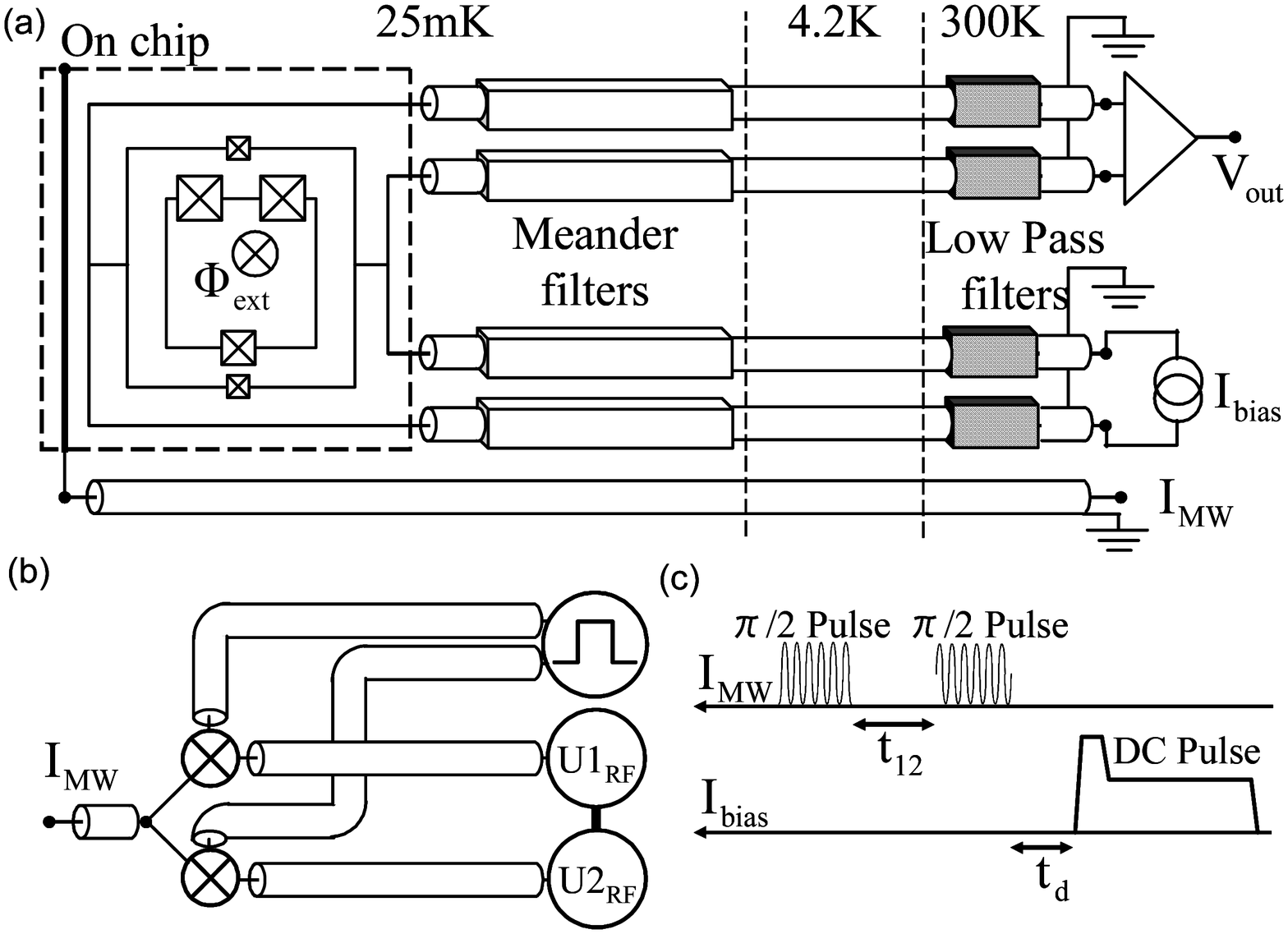}
\caption{\label{fig:sample} 
\footnotesize
(a)	 Schematic drawings of the measurement circuit.
 The squares with crosses represent Josephson junctions.
 Microwaves (MW) were applied to the qubit from the on-chip strip line.
(b)	Schematic drawings of the phase-shifted double pulse generation.
The relative phase shift between two pulses was precisely controlled with two synchronous MW generators $U1_{RF}$ and $U2_{RF}$, each MW-pulse was re-shaped by multiplying with a recutangular pulse from a synchronous driven pulse generator, then a pair of phase-controlled MW-pulses were delivered to the sample through a microwave adder. 
(c)	Timing chart of the resonant $\frac{\pi}{2}$ MW-pulses separated by time $t_{12}$ and a readout DC-pulse. 
}
\end{figure}

 Figure 1(a) shows schematic drawings of the measurement circuit; a superconducting qubit (an inside loop with three Josephson junctions of critical current $I_{\rm c}^{\rm qubit}\simeq 0.6 \mu$A for larger junctions) and an under-damped dc-SQUID (an outside loop with two small Josephson junctions of critical current $I_{\rm c}^{\rm SQ}\simeq 0.15 \mu$A for each junction) as a detector.
 The qubit contains three Josephson junctions, two of which have the same Josephson coupling energy $E_{\rm J}=\hbar I^{\rm qubit}_{\rm c}/2e$.
 The third has ${\alpha}E_{\rm J}$, with $0.5<\alpha<{1}$. The $\alpha$ value can be controlled by designing the ratio of the area of the smallest Josephson junction to the other two larger junctions in the qubit.
 We used a sample with $\alpha\simeq 0.7$ and the areas of the larger and smaller junctions were 0.1$\times$0.3 $\mu$m$^{2}$ and 0.1$\times$0.2 $\mu$m$^{2}$, respectively.
 The loop area ratio of the qubit to the SQUID was about 3:5.
 The dc-SQUID with two Josephson junctions was under-damped with no shunt resistor.
 The qubit and the dc-SQUID were coupled magnetically via mutual inductance $M\simeq7$ pH.
 The aluminum Josephson junctions were fabricated by using suspended bridges and the shadow evaporation techniques  \cite{D-APL77}.
 By carefully designing the junction parameters   \cite{MOLTWL-Sci99,Orando_PRB,Wal}, the inner loop can be made to behave as an effective two-state system   \cite{Saito,Nakano}.
 In fact, the readout result of the qubit changes greatly with the qubit design, ranging from the purely classical to the quantum regime  \cite{Takayanagi,Tanaka}.

\begin{figure}
\includegraphics[width=9cm]{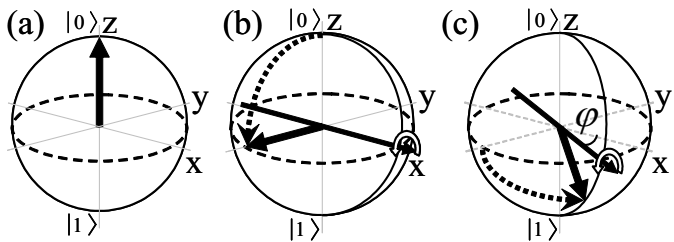}
\caption{\label{fig:sample}
\footnotesize
 Schematic diagram of qubit vector motion induced by the phase-shifted double $\frac{\pi}{2}$ on-resonance pulses ($\omega=\omega_{0}$). It is described in the rotating frame of the qubit Larmor frequency $\omega_{0}$. (a) The qubit vector in the initial state. The qubit is in the ground state $|0\rangle$.  (b) The first resonant $\frac{\pi}{2}$ pulse ($\varphi=0$) tips the qubit vector to the equator. The qubit vector remains there, because the on-resonance pulse is used. (c) The second resonant $\frac{\pi}{2}$ pulse, in which the phase-shift $\varphi \neq 0$ is introduced, tips the qubit vector on another axis, which is at an angle $\varphi$ from the $x$-axis.}
\end{figure}

 The sample was cooled to 25 mK with a dilution refrigerator and it underwent a superconducting transition at $\sim$1.2 K.
 In order to reduce external magnetic field fluctuations, both the sample holder and the operating magnet were mounted inside a three-fold $\mu$-metal can.
 As schematically shown in Fig.1(a), the qubit is biased with a static magnetic flux $\Phi_{\rm ext}$ using an external coil.
 A microwave on-chip strip line was placed at 60 $\mu$m from the qubit to induce oscillating magnetic fields in the qubit loop.
 The switching voltage of the SQUID was measured by the four-probe method.
 The generation of phase-shifted operating pulses is shown schematically in Fig. 1(b). The relative phase shift between two pulses was precisely controlled with two synchronous MW generators $U1_{RF}$ and $U2_{RF}$. In order to trim the pulse shape and reduce the noise level, each pulse-modulated MW burst will be re-shaped by multiplying it with a rectangular pulse from a synchronous driven pulse generator, then a pair of phase-controlled pulses are delivered to the sample through a microwave adder. Figure 1(c) shows the timing chart of the operating pulses and the readout pulse. The on-chip MW line provides a microwave current burst which induces an oscillating magnetic field in the qubit loop. We adjusted the timing of a readout DC-pulse, which is delivered to the detector SQUID through the current bias line ($I_{\rm bias}$) just after the second $\frac{\pi}{2}$ control pulse. 
The width and amplitude of the $\frac{\pi}{2}$ pulse are determined by the Rabi oscillation at the resonant frequency($\omega_{0}$). The relative phase shift between these two pulses was precisely controlled with two synchronous microwave generators. The bias current line delivers the readout pulses, and the switching event is detected on the signal line ($V_{\rm out}$). As regards the shape of the bias pulse, its peak width of 150 ns, is limited by the rise time of the filters installed in the bias current line. The width of the trailing plateau of 600 ns, and the trailing height ratio of 0.7, were selected in order to optimize the discrimination of the switching voltage of the dc-SQUID detector \cite{CNHM-Sci03}.

\begin{figure}
\includegraphics[width=9cm]{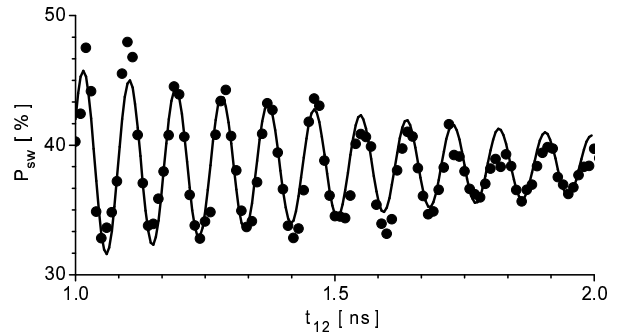}
\vspace*{-10mm}
\caption{\label{fig:sample}
\footnotesize
 On-resonance Ramsey fringes observed by using the phase-shifted double $\frac{\pi}{2}$ pulse technique.
 The Larmor frequency is $\omega/2\pi$=11.4 GHz.
 The width of the $\frac{\pi}{2}$ pulse, 5 ns, is determined by Rabi oscillation.
An exponentially damped sinusoidal curve fitted with the decay time constant $T_{2} =$ 0.84 ns is also shown. }
\end{figure}

 The qubit is described by the Hamiltonian $\hat{H_{0}}=\frac{\hbar}{2}(\varepsilon_{f} \hat{\sigma}_{z}+\Delta \hat{\sigma}_{x}),$ where $ \hat{\sigma}_{x,z}$ are the Pauli matrices. We estimated the qubit tunnel splitting $\frac{\Delta}{2\pi}\simeq 1$ GHz from the spectroscopy. 
 The two eigenstates of $\hat{\sigma}_{z}$ are macroscopically distinct states with the qubit supercurrent circulating in opposite directions, i.e., the clockwise state $|R\rangle$ and the counter-clockwise state $|L\rangle$.
 The dc-SQUID picks up a signal that is proportional to $\hat{\sigma}_{z}$.
 To control the qubit state we employ a microwave current burst, which induces oscillating magnetic fields in the qubit loop.
 That event is described by the perturbation Hamiltonian $\hat{V}=\frac{\hbar}{2} a(t)\hat{\sigma}_{z}$, where $a(t)=a\cos(\omega t + \varphi)$ has an amplitude correlated to the power of the applied microwave pulse, $\omega$ is the angular frequency and $\varphi$ represents the origin of the phase of the applied microwave pulse.
To maintain the analogy between the spin $\frac{1}{2}$ and the flux qubit, we write the Hamiltonian in the rotating frame approximation, $\hat{H}_{\rm rot}=\frac{\hbar\omega}{2} \hat{\sigma}_{z} + e^{-i\frac{\omega t}{2} \hat{\sigma}_{z}} (\hat{H_{0}}+\hat{V}) e^{i\frac{\omega t}{2} \hat{\sigma}_{z}} $,  in terms of the energy eigenstates ($|0\rangle$, $|1\rangle$) and obtain

\begin{equation}
\hat{H}_{\rm rot}=\frac{\hbar}{2} 
\left( 
\begin{array}{cc}
\omega-\omega_{0} & -a e^{i{\varphi}} \sin{\theta}\\
-a e^{-i{\varphi}} \sin{\theta} & \omega_{0}-\omega\\
\end{array}
\right)
\end{equation}

\noindent
where $\omega_{0}=\frac{1}{\hbar}\sqrt{{\varepsilon_{f}}^2+\Delta^2}$ is the qubit Larmor frequency at the measured flux bias and $\theta \equiv \arctan{\frac{\Delta}{\varepsilon_{f}}}$. 

 Figure 2 shows schematic diagrams describing how the qubit vector is operated during the Ramsey fringe experiment with the phase shift technique in a rotating frame.
 We assume that the initial state of the qubit is the ground state $|0\rangle$. The first resonant $\frac{\pi}{2}$ pulse ($\varphi=0$) tips the qubit vector towards the equator with the $x$-axis as the rotating axis ($\hat{H}_{\rm rot}\propto \hat{\sigma}_{x}$). The qubit vector remains there because we introduce no detuning ($\omega=\omega_{0}$).  After a time $t_{12}$, the second resonant $\frac{\pi}{2}$ pulse with a given phase shift $\varphi\neq 0$ tips the qubit vector on another axis at an angle $\varphi$ from the $x$-axis. The resulting qubit vector does not reach the south pole ($|1\rangle$) of the Bloch sphere. The detector SQUID switches by picking up the $z$-component of the final qubit vector after the trigger readout pulse. Repeating this sequence typically 10,000 times, with a fixed $t_{12}$, we obtain the switching probability. Figure 3 shows the damped sinusoidal oscillation obtained by changing the pulse interval $t_{12}$.  The phase shift of the second pulse was programmed from the following relation ; $\varphi=\omega_{0} t_{12}$ mod $2\pi$. This equation gives a 2$\pi$ phase change to the resonant microwave pulse during a period of $T=\frac{2\pi}{\omega_{0}}$. This means that we introduce a phase shift with the Lamor frequency. 
 
When Ramsey fringes are observed in the conventional way, a few hundred MHz detuning is typically introduced near the qubit Larmor frequency, i.e., $\sim 100$ MHz detuning at a Larmor frequency of $\sim 5$ GHz. With this detuning method, after the first detuned $\frac{\pi}{2}$ pulse, the qubit vector rotates along the equator of the Bloch sphere with this detuning frequency, $\sim 100$ MHz \cite{CNHM-Sci03}. If we use this method to control the qubit azimuth angle, a time of $\sim 10$ ns is required for every $2\pi$ azimuth angle rotation of the qubit vector. This operating time cannot be as short as 1 ns, because a detuning of 1 GHz does not work properly.
However, with the phase shift technique with the resonant frequency, as we have shown, it is possible to revolve the rotational axis of the qubit vector within the $xy$-plane with the frequency above 11 GHz. Using the following relation $Z(\phi)=X(\frac{\pi}{2})Y(\phi)X(-\frac{\pi}{2})$, the azimuth angle $\phi$ rotation on the $z$-axis can be decomposed into three successive rotational operations such that $-\frac{\pi}{2}$ rotation on the $x$-axis, $\phi$ rotation on the $y$-axis, and $\frac{\pi}{2}$ rotation on the $x$-axis. If the qubit is driven strongly enough, each $\frac{\pi}{2}$-pulse width can be as short as 0.1 ns, therefore the total composite operation $X(\frac{\pi}{2})Y(\phi)X(-\frac{\pi}{2})$ can be completed in $\sim 1$ ns.

Compared with the conventional detuning method, the phase shift technique provides us with the opportunity to increase the speed of the qubit unitary gate operation by more than an order of magnitude. This method will save operating time and we can make best use of the precious coherence time. 

We thank M. Ueda, J. E. Mooij, C. J. P. M. Harmans, Y. Nakamura, I. Chiorescu, D. Esteve, D. Vion, for useful discussions. This work was supported by the CREST project of the Japan Science and Technology Agency (JST). 


\end{document}